%% file: V2_with_name.tex
\documentclass[runningheads]{llncs}
\usepackage{graphicx}
\usepackage{color, colortbl}
\usepackage{stfloats}
\usepackage[most]{tcolorbox}
\usepackage{url}
\usepackage{xspace}
\usepackage{todonotes}
\usepackage{amsfonts} 
\usepackage{tabularx}
\usepackage[inline,shortlabels]{enumitem}
\usepackage[shortcuts]{extdash}
\usepackage{booktabs}
\usepackage[english]{babel}
\usepackage{blindtext}
\usepackage{multirow} 
\usepackage{subcaption}
\usepackage{threeparttable}
\usepackage{placeins}
\usepackage{lscape}
\usepackage[utf8]{inputenc}
\usepackage{hyperref}
\usepackage{float}

\renewcommand{\texttt}[1]{{\vspace{0pt}\tt#1}}

\usepackage{hyphenat}
\usepackage[shortcuts]{extdash}
\newcommand{\clvr}{CLVR\xspace}

\newcommand{\newpar}[1]{\vspace{0.05cm}\noindent\textbf{#1.}\xspace}

\newcommand{\GSR}[0]{GSR\xspace}
\newcommand{\VHGSR}[0]{VHGSR\xspace}

\begin{document}
\title{CLVR Ordering of Transactions on AMMs}



%
\titlerunning{\clvr Ordering of Transactions}
%

\author{Robert McLaughlin\inst{1} \and
 Nir Chemaya\inst{2} \and
 Dingyue Liu\inst{3} \and
 Dahlia Malkhi \inst{4} \inst{5}}

\institute{
 Sigma Prime \and
 Department of Economics, Ben Gurion University of the Negev \and
 Uniswap Labs \and
 Department of Computer Science, University of California, Santa Barbara \and
 Chainlink Labs
 }


%

%

%
\maketitle              
\begin{abstract}
This paper introduces a trade ordering rule that aims to reduce intra-block price volatility in Automated Market Maker (AMM) powered decentralized exchanges. 
The ordering rule introduced here, Clever Look-ahead Volatility Reduction (\clvr), operates under the (common) framework in decentralized finance that allows some entities to observe trade requests before they are settled, assemble them into ``blocks'', and order them as they like. 
On AMM exchanges, asset prices are continuously and transparently updated as a result of each trade and therefore, transaction order has high financial value. 
\clvr aims to order transactions for traders' benefit. 
Our primary focus is intra-block price stability (minimizing volatility), which has two main benefits for traders: it reduces transaction failure rate and allows traders to receive closer prices to the reference price at which they submit their transactions accordingly.
We show that \clvr constructs an ordering which approximately minimizes price volatility with a small computation cost and can be trivially verified externally.



\end{abstract}
%


\input{sections/01.Introduction}

\input{sections/02.Background}
\input{sections/03.Model}
\input{sections/04.Optimality}

\input{sections/05.Simulation}

\input{sections/06.Conclusions}
\clearpage

%
%
\bibliographystyle{splncs04}
\bibliography{cas-refs}

\clearpage

\appendix

\input{sections/XX.Appendix}

\end{document}

%% file: sections/01.Introduction.tex
\section{Introduction}
A new financial market has surfaced in the past decade on Ethereum and other blockchain platforms.
The financial activities on these platforms, referred to as Decentralized Finance (``DeFi''), represents around \$1T value~\cite{defillamadefi} and completely changes the rules of the financial ``game'': (i) 
settlement on these platforms allows for transparent and continuous price update, and (ii) the platform itself can see and (re)order trades.

Today, (re)ordering is primarily used for selfish profit extraction \cite{daian2020flash,kulkarni2022towards}, and consequently, considerable effort has been directed towards preventing reordering. 
\cite{chitra2023towards} suggests that sequencing rules to prevent unfair reordering in blockchains should be application-specific, such as tailored rules for AMMs.
For example , Ferreira, Parkes, and Parkes, 2023 \cite{xavier2023credible},  offer an ordering rule tailored for AMMs known as the Greedy Sequencing Rule (\GSR), specifically designed to mitigate manipulation and ``sandwich attacks''. GSR is designed with blockchain readiness in mind. By being easy to verify, agents could monitor GSR's enforcement and ensure that block builders follow this rule.

This paper is aligned with the \GSR philosophy: establish simple and easy to implement ordering rules to provide financial protection in DeFi. However, 
rather than asking how to prevent predatory reordering, our work is focused on generally benefiting traders and the DeFi economy. It asks the following question: suppose that a ``social planner'' could order transactions, what would this order aim to achieve?\footnote{It is important to note that the focus is on the order itself, not mechanisms for enacting it, which are orthogonal to this work.}

To take a first stab at this question, like \GSR we narrow the exploration to transaction ordering within one popular form of DeFi, a decentralized Automated Market Maker (AMM)~\cite{werner2021sok}, which allows users to swap digital assets (tokens) on the blockchain in a trustless setting.
An AMM is a smart contract that enables users to swap one token for another at a price that is determined fully automatically by the smart contract itself.
Unlike other trading mechanisms, AMMs do not match trading parties with counterparties, but instead custody and oversee a ``liquidity pool'' of reserve assets, which they use to service these swaps. 
An AMM's price updates continuously as swaps are executed -- put simply, prices rise with `buy' orders, and fall with `sell' orders.
Consequently, each transaction's effective execution price is influenced by the transactions that precede it.

AMMs are extremely popular, with daily trading volume reaching over \$1 billion~\cite{defillamadex}.
Furthermore, the impact of AMMs is extending beyond the DeFi ecosystem.
Extensive discussion of the advantages of adopting this new trading innovation has emerged in traditional financial markets, such as stock markets and foreign exchanges~\cite{adams2023chain,BIS2023,malinova2023learning}. 

The key idea in this work is to examine the impact of ordering on \textbf{price stability} (volatility) on AMMs.
Our focus on this measure stems from its particular suitability as an economic goal for blockchain systems.
In particular, price stability has two essential aspects that traders benefit from:

\begin{itemize}
    \item First, traders set their \emph{slippage tolerance} (a minimum of what they are willing to accept for their trade, i.e., a price limit). When the execution price of a transaction yields the trader less than the specified minimum, the transaction will fail, but the trader will still incur gas fees. This results in a double economic loss for the trader: they must pay gas fees without execution and cannot capture any financial benefit of the trade. Therefore, reducing price volatility can help lower the chance of such failures.

    \item Second, when placing an order traders typically consult the pool's current price as a reference when making trading decisions. The current price of a liquidity pool is based on the smart contract's state in the most recent block --- traders typically observe this price through an interface like the Uniswap website\footnote{Uniswap is one of the largest AMMs in DeFi \cite{lehar2021decentralized}}. Traders expect their transaction's actual execution price to be not much different than the reference price, even though it may change arbitrarily due to market conditions.
\end{itemize}

In order to protect price stability, we introduce Clever Look-ahead Volatility Reduction (\clvr), a practical method for computing (nearly) optimal transaction ordering for minimizing volatility.
\clvr approximates the maximal price stability (minimal volatility) for a batch of $n$ transactions with a polynomial-time computational cost ($O(n^2)$).
The key idea of \clvr is picking at each step as next trade the one that minimizes local one-step price-deviation from a starting status quo price. Despite ``only'' approximating the optimal price-stability, \clvr exhibits several desirable properties: 

\begin{enumerate}
    \item We prove that \clvr protects traders from traditional three-transaction sandwich attacks.
    \item We demonstrate the cases under which \clvr achieves optimal price stability, and provide evidence experimentally that \clvr attains approximately optimal results.
    \item We show that \clvr achieves superior price stability compared to \GSR.
    \item We show that \clvr reduces the transaction failure rate relative to a random ordering.
\end{enumerate}

We evaluate the impact of transaction ordering using a simulation over synthetic workloads extrapolated from empirical data with a variety of parameters like block sizes, liquidity availability, and transaction sizes.
Crucially, we find that when traders use slippage tolerance settings, CLVR transaction ordering can significantly reduce traders’ transaction failure rates by approximately 95\% compared to random ordering.

We also examine the block size itself, which represents the duration that the social planner sets for trading frequency on an AMM.
We investigate whether, if given more time (a larger block size), the planner can order trades to meet a price stability objective more effectively.
This evaluation reveals the following result: we show that larger blocks give ordering algorithm such as \clvr more flexibility to organize trades and improve price stability. Section \ref{Empirical-Data} demonstrates 85\% reduction in price volatility on empirical data.

Price volatility can be further reduced when traders split their transactions under an ordering algorithm.
We show that when using \clvr, traders fare better when they break down trades into smaller orders (which our system can sequence between other orders). More specifically, we show that transaction splitting can dramatically improve price stability due to increased flexibility available to the \clvr ordering rule. 
An ordering algorithm that encourages traders to split their transactions into smaller ones could substantially impact price stability, underscoring the importance of this decision.
We show that under \clvr, even in a small block with only two transactions, traders can enjoy a reduction of price volatility of 6 orders of magnitude just from splitting their transaction into small ones. 
In section \ref{sec:Splitting}, we test whether traders can benefit from splitting their transactions under \clvr, and we find that all traders do indeed gain from doing so.
Finally, Section~\ref{subsec:worst} explores \clvr’s worst-case performance and provides an upper-bound estimate of approximately an 8\% deviation from optimality. It also explains the conditions under which \clvr achieves optimal price volatility and identifies the settings that affect price volatility.

In summary, our work highlights the importance of transaction ordering, beyond the scope of mitigating MEV and price manipulation, in terms of a public good that improves the price stability.
It provides evidence that even today, when the number of transactions executed in a given block in an AMM are still relatively low, ordering transactions can help improve price stability.
If an AMM is to scale up to the traffic observed in traditional financial markets, then good ordering mechanisms are necessary because ordering matters, and has a large impact on the economy that can be leveraged in a positive way.

%% file: sections/02.Background.tex
\section{Background and Related Works}
\label{sec:Background}

\paragraph{The Transaction Ordering Issue.}

When trading with an AMM, traders interact directly with a market-making smart contract in an order determined by the host blockchain system (e.g., Ethereum).
This has  implications on many aspects of consumer protection. 

Unlike traditional markets, ordering transactions on blockchains like Ethereum is carried in batches, known as blocks.
Each block is produced by a single leader in the network, and the network rotates the leader every block.
The block producer is given exclusive authority to sequence transactions within the block that they construct, and thus, are able to insert, re-order, and/or censor transactions, as mentioned above.

In contrast, in traditional markets, execution ordering mechanisms are often enforced by regulatory bodies (typically, on a first-come-first-served basis)~\cite{harvey2024evolution}.
This is difficult to enforce within a distributed system.
Indeed, the concept of ordering events by time in distributed systems is subtle to define and enforce (e.g.,~\cite{cryptoeprint:2021/139,cachin2022quickorderfairness,kelkar2023themis,zhang2020byzantine}).
Currently the Ethereum blockchain makes no attempt at enforcing ordering fairness, and instead gives the block-builder a monopoly power that they can use to order the transactions as they wish.


The ability to insert, re-order, and/or censor transactions is thus highly \emph{financially} valuable: one can use this power to both manipulate users' swap prices, and to exploit opportune prices.
An example of the former is a \emph{sandwich attack}, which involves extracting profit by forcing a user to swap at a price that was manipulated to be higher than expected~\cite{park2023conceptual,heimbach2022sok,chemaya2023power}.
An example of the latter is \emph{arbitrage}, which involves buying an asset for a low price in one market, and selling it for a higher price in another market, yielding a profit~\cite{lehar2023battle,mclaughlin2023large}.
Using this ability to extract value from the network is extremely common, and is conceptually known as Maximum Extractable Value (MEV)~\cite{daian2020flash}.\footnote{Originally, the term MEV referred to Miner Extractable Value\cite{daian2020flash}.}
MEV extraction requires specialized knowledge and tooling, so block producers often outsource block production to a third-party via MEV-Boost, an auction system whereby third-parties bid for sequencing rights~\cite{capponi2024proposer}.

\paragraph{Existing Mitigation Strategies.}

\cite{daian2020flash,park2023conceptual,lehar2023battle}  introduce the economic value of ordering transactions in an AMM (MEV).
There has been significant recent research regarding the impact of MEV extraction, both in terms of the welfare of blockchain users and the technical operation of the blockchain itself~\cite{yang2022sok,capponi2023maximal}.
One thread of research seeks to address the critical question of transaction ordering by proposing ``fair ordering'' policies that blockchain protocols can use to mitigate MEV activity~\cite{yang2022sok,xavier2023credible}.
However, imposing those policies on a blockchain can be challenging, and does not always achieve the desired outcome; for example, \cite{oz2024study} shows the presence of MEV extraction techniques even in first-come, first-serve transaction ordering blockchains. In addition, it is challenging to design ordering rules that improve outcomes across all DeFi and blockchain applications, so \cite{chitra2023towards} suggest they should be application-specific.

Ferreira, Parkes, and Parkes, 2023 \cite{xavier2023credible} were the first to offer an ordering mechanism specifically designed for AMMs, known as the Greedy Sequencing Rule (\GSR).
Their first goal is to reduce unnecessary price deviation from a block's initial price and to prevent sandwich attacks; an additional, pragmatic concern was to develop an ordering rule that would be ``credible'', i.e., easy to verify, so that agents could monitor its operation and ensure that block builders follow this rule. In the \GSR, when a buy/sell transaction is executed and causes the liquidity pool price to deviate from the initial price, the algorithm subsequently picks a transaction in the ``opposite direction'' (if one exists), which ensures that the price does not continue diverting to extremes (when possible).
If the subsequent price movement pushes the price beyond the initial price, the algorithm will again alternate from buy(sell) to sell(buy).
\cite{xavier2023credible} demonstrate that this strategy eliminates sandwich attacks.
In addition, \GSR has a low computational cost: it only needs to classify pending transactions as `buy' or `sell', and pick ones trading in the right direction.
This low computational cost and straightforward design allows anyone to monitor and easily verify compliance with the sequencing rule.

\GSR is further enhanced in \cite{ankile2023see}, where they offer the Volume Heuristic Greedy Sequencing Rule (\VHGSR), which, on top of the \GSR method (namely, alternate between buy and sell), the algorithm prioritizes selecting small transactions before bigger ones.
They also show that \VHGSR can give a reasonable approximately optimal price volatility for a small set of pending transactions, $n<8$.
With respect to price volatility however, for some pending transactions (especially in big blocks), the optimal volatility-minimization order of transactions violates the \GSR and \VHGSR, as demonstrated in section ~\ref{subsec:Performance}.

Another way is to protect traders by setting a slippage tolerance, 
which sets a limit price above which a trade will not execute~\cite{chemaya2023power,heimbach2022sok}.
While this approach can protect traders from price manipulation and sandwich attacks, it does not
contemplate the economic good offered by imposing an ordering mechanism, as we offer in this work.

An alternative way to protect users is to generate an auction between the traders to determine priority of their transactions~\cite{ernst2022would}.
Still, this approach will not leverage the main benefit of ordering a transaction to optimally achieve economic goals such as reducing price deviation or inequality.
One way to address this issue is to offer batch auctions, which first intended to mitigate high-frequency trading on traditional markets\cite{budish2015high}. 
The main idea of this approach is to collect all pending transactions and run an auction to determine a single price that clears the market in a way that does not give any particular trader a better exchange price relative to another. \cite{ramseyer2022augmenting,zhangcomputation,canidio2023arbitrageurs} offer models to implement this idea in AMMs.
Our approach is similar to the batch auction in that it leverages the idea of executing transactions together to clear the market better and provide better prices.
However, we offer to achieve this goal on an AMM by implementing an ordering mechanism that can be used without modifying the particular mechanics of how an order is filled.

Another way is to adopt a norm of ordering transactions using a credible rule that is easy to monitor and find a way to punish any block producer that deviates from this norm.
As mentioned above, such rules were introduced, e.g., in \GSR~\cite{xavier2023credible,li2023mev} to protect traders from sandwich attacks and provide better exchange prices for traders.

Finally, previous work has explored goals other than price stability in AMMs, such as comparing them to centralized exchanges and limit order book mechanisms, and examining the risks and benefits of liquidity provision in these markets, and more~\cite{aoyagi2021coexisting,lehar2021decentralized,harvey2024evolution,exchanges2022quality,milionis2022automated,hasbrouck2022need,capponi2021adoption,milionis2023effect,milionis2023myersonian,angeris2020improved,angeris2022does}.
Our main contribution to this area of research is to shed light on how an AMM can improve price stability while imposing a transaction ordering mechanism.

%% file: sections/03.Model.tex
\section{Model}
\label{sec:model}
There are several different AMM designs.
For this work we will focus only on the Uniswap V2~\cite{adams2020uniswap} AMM mechanism (known as a ``Constant-Product Market Maker'', or CPMM), as it is both popular and relatively simple.
A CPMM's pricing rule is defined as follows:
$k = x_t y_t$, where $k$ represents a \textit{potential} which remains constant, and $x_t$ and $y_t$ represent the quantity of tokens X and Y in the AMM's liquidity pool at time $t$.
The current spot price of the pool is simply the ratio of tokens held in the AMM's reserve at time $t$, computed as $p_t=\frac{y_t}{x_t}$.

Without loss of generality, when a trader wishes to swap token X for token Y at time $t$, the trader will add $\Delta x_t$ tokens X to the pool, and withdraw $\Delta y_{\textup{out},t}$ tokens Y from the pool, according to Equation~\ref{eq:general_invariant}.
\begin{align}
k &= (x_t + \Delta x_t (1 - f))(y_t - \Delta y_{\textup{out},t}) \label{eq:general_invariant} \\
k &= (x_t + \Delta x_t)(y_t - \Delta y_{\textup{out},t}) \label{eq:frictionless_invariant}
\end{align}
Where $f$ is the \emph{fee rate} (set to $0.003$ on Uniswap V2).
For simplicity, we consider a frictionless trading environment with $f = 0.0$, expressed in Equation~\ref{eq:frictionless_invariant}.

Prices update continuously as trades are processed.
In this work, we examine several methods of ordering trades, using the following notation.
The set of available trades is denoted as $M = \left\{\alpha:~\textup{sell}~a,~\beta:~\textup{buy}~b,~\ldots\right\}$, where each trade indicates a direction (sell or buy) and a quantity ($a$ or $b$), representing the amount of token the trader trades.
We refer to trades by their \emph{labels} ($\alpha$, $\beta$, etc.) for convenience.
Transactions that `sell' send some quantity of X tokens to the AMM and receive a quantity of Y tokens; transactions that `buy' do the opposite.

We notate a particular sequence (ordering) of trades with the bijection $\sigma: \{1,~\ldots,~n\} \rightarrow \{\alpha,~\beta,~\ldots \}$, which maps a position in the sequence to a specific trade.
In Appendix~\ref{appendix:model_equations} we derive functions $x_\textup{out}(\sigma,~i)$ and $y_\textup{out}(\sigma,~i)$, which indicate the amount received by the $i^\textup{th}$ trade in ordering $\sigma$.
We also define $X(\sigma,~i)$ and $Y(\sigma,~i)$, which indicate the pool's reserves of tokens X and Y after executing the $i^\textup{th}$ trade in ordering $\sigma$.

Finally, the spot price updates as follows:
\begin{align}
    P(\sigma, i) &= \frac{Y(\sigma, i)}{X(\sigma, i)} \label{eq:price_update}
\end{align}

Equation~\ref{eq:price_update} shows that the price changes continuously as trades are processed, meaning the order of transactions directly affects the exchange price for each trade.
%
%
The following example illustrates a situation where transaction ordering significantly impacts execution price.
Let $x_0 = y_0 = 100$, then $k = x_0 y_0 = 10,000$.
Now, consider a set of two pending transactions $\left\{\alpha:~\textup{sell}~10,~\beta:~\textup{sell}~10\right\}$, which we order as $\sigma:~\{1 \mapsto \alpha,~2\mapsto \beta\}$.
By applying Equation~\ref{eq:frictionless_invariant}, we compute the amount of token received for each transaction: $y_{\textup{out}}(\sigma,1)=9.09$, $y_{\textup{out}}(\sigma,2)=7.58$.
Observe that despite each trader spending the same quantity of token X (10), they each receive significantly different amounts of token Y, depending on execution order.

In the above example, execution ordering impacts which trader receives $9.09$ units of Y, and which trader receives $7.58$ units of Y.
However, ordering a larger number of transactions involves a much more complex set of price decisions.
Consider, for example, a sequence of four pending transactions: $\left\{\alpha:~\textup{buy}~10,\right.\allowbreak~\beta:~\textup{buy}~10,\allowbreak~\gamma:~\textup{sell}~10,\left.\allowbreak~\delta:~\textup{sell}~10\right\}$.
Let's consider two possible orderings: $\sigma_1$ process all sell orders then process all buy orders, and $\sigma_2$ alternates between processing `buy' and `sell' transactions.
In ordering $\sigma_1$ we compute the sequence of outputs: $y_{\textup{out}}(\sigma_1,1)=9.09$, $y_{\textup{out}}(\sigma_1,2)=7.58$, $x_{\textup{out}}(\sigma_1,3)=12.86$, and $x_{\textup{out}}(\sigma_1,4)=10.37$.
However, in ordering $\sigma_2$ we compute the sequence of outputs: $y_{\textup{out}}(\sigma_2,1)=9.09$, $x_{\textup{out}}(\sigma_2,2)=10.90$, $y_{\textup{out}}(\sigma_2,3)=9.25$, and $x_{\textup{out}}(\sigma_2,4)=10.73$.
These two sequences demonstrate outcomes that differ not just in which execution price is awarded to whom, but it also significantly impacts the execution price levels themselves.
In Section~\ref{sec:optimality} we discuss the direct implications that this has on fairness, inequality, and price volatility.


%% file: sections/04.Optimality.tex
\section{Optimal Ordering of Transactions}
\label{sec:optimality}

In this section, we introduce an economic measure against which we will seek to optimize transaction ordering.
An intuitive way to think about this is to imagine that a social planner gets a set of pending transactions and an objective function that it needs to optimize.
The social planner's role is only to pick the order of the transactions, and they cannot censor or insert transactions.
The following section 
offers one such measure, price stability, for which social planners may decide to optimize.
However, finding the optimal ordering can be costly.
In Section~\ref{subsec:Computation cost} we describe our mechanism, \clvr, that provides a fast method to approximate a good transaction ordering for price stability.


\subsection{Price Stability}
\label{subsec:Price Stability}
For a given set of pending transactions $M$, the ordering of transactions $\sigma$ that will give the economy optimal price stability is the one that minimizes the \emph{intra-block volatility} of traders' exchange prices from the ``status-quo'' price.
In Ethereum today, this \emph{intra}-block volatility is solely under the control of the sequencer. Later, we control for volatility by proposing an algorithm which the block sequencer must follow.

It is important to mention here that our definition of volatility differs slightly from that in traditional markets.
Traditionally, volatility is a measure of the price movement \emph{over a period of time}.
However, in DeFi, trades that are executed within the same batch (block) are effectively instantaneous to each other.
As such, our measure of volatility imagines a sort of ``fictitious time'' in order to capture a metric which reflects the pricing action that occurs when executing trades one-by-one within a block.
An important consequence of this is that we are not attempting to control \emph{inter-}block volatility (as seen evolving in actual time), which is influenced by information like news and other price-discovery dynamics.
In fact, in our model the opening and closing prices of a block are not (very) sensitive to the ordering chosen within, thus leaving \emph{inter-}block volatility nearly unchanged -- this can be seen visually in Figures~\ref{fig:ex1} and~\ref{greedy}.

\medspace

We set the ``status-quo'' price to be the opening price, before any trades in the set of pending trades $M$ begin to execute.
We expect that traders consider this price to be fair, as they were able to observe and consider this price before broadcasting their trade.
In addition, most AMM protocols (Uniswap, Pancakeswap, etc.) set the slippage tolerance according to this opening price.
Maintaining price stability will attempt to execute more orders around the status-quo price, and will thus reduce the number of unnecessary transaction failures.
However, other potential ways exist to set the ``status-quo'' price as described in Section~\ref{sec:conclusion}.
In this case, the price volatility for a particular ordering $\sigma$ can be captured by using the following function:
\begin{align}
\textsc{Vol}(\sigma)= \frac{1}{n} \sum_{i=1}^{n}  \left(\ln~p_0~-~\ln~P(\sigma,~i) \right)^2
\end{align}
Where $n = \vert M \vert$ (the fictitious time period) and $p_0$ is the ``status-quo'' spot price. 
This function is based on a standard price volatility measure from \cite{liu1999statistical} adapted to the AMM setting.\footnote{\cite{liu1999statistical} defines volatility as the average change in the logarithm of prices. While this paper focuses on the prices of stock indices and their fluctuations, we use this approach for AMMs, where volatility is measured relative to the initial price of the liquidity pools rather than the fluctuations of assets over time.}
Also note that this measure is insensitive to the choice of spot price direction -- i.e., whether $p_0 = x_0 / y_0$ or $p_0 = y_0 / x_0$.
A brief proof is provided below. 
In order to achieve maximal price stability the social planner will need to \emph{minimize} this function over all such possible transaction orderings $\sigma$.

\newpar{Proof}
Below is a brief proof of the insensitivity to quote direction.
In a nutshell, this is equivalent to choosing a unit of account, i.e., whether to quote the USD-JPY exchange rate in dollars per yen, or yen per dollar.
This becomes important when measuring how much the price has deviated from the starting point.
A na\"ive approach would be to simply compute the absolute difference, however this makes the computation sensitive to which way the spot price is quoted.
Put explicitly,

\begin{align}
\label{eq:naive}
\left( \frac{\alpha_0}{\beta_0} - \frac{\alpha_0 + \Delta \alpha}{\beta_0 - \Delta \beta} \right)^2 \not= \left( \frac{\beta_0}{\alpha_0} - \frac{\beta_0 - \Delta \beta}{\alpha_0 + \Delta \alpha} \right)^2
\end{align}

\clvr's computation is insensitive to this choice, as it considers instead the absolute difference in \emph{log} prices, thus making the computation insensitive to the quote direction.
We prove this statement below.

\begin{align}
\left( \ln \frac{\alpha_0}{\beta_0} - \ln \frac{\alpha_0 + \Delta \alpha}{\beta_0 + \Delta \beta} \right)^2 &= \left( - \left( \ln \frac{\alpha_0}{\beta_0} - \ln \frac{\alpha_0 + \Delta \alpha}{\beta_0 + \Delta \beta} \right) \right)^2 \\
&= \left( \left( - \ln \frac{\alpha_0}{\beta_0} \right) - \left(-\ln \frac{\alpha_0 + \Delta \alpha}{\beta_0 + \Delta \beta} \right)\right)^2\\
&= \left( \ln \left[\left(\frac{\alpha_0}{\beta_0} \right)^{-1}\right] - \ln \left[\left(\frac{\alpha_0 + \Delta \alpha}{\beta_0 + \Delta \beta} \right)^{-1}\right]\right)^2\\
&= \left( \ln \frac{\beta_0}{\alpha_0} - \ln \frac{\beta_0 + \Delta \beta}{\alpha_0 + \Delta \alpha} \right)^2 \quad
\end{align}

\subsection{Approximating and Verifying Optimal Ordering for Price Stability}
\label{subsec:Computation cost}

While ideally, the social planner will pick the optimal ordering of transactions with the best price stability, the computational cost of exhaustively going through all permutations of $n$ transactions is prohibitive: 
even at one billion evaluations carried per second, it will take one hundred hours to exhaustively scan all orderings for (say) $17$ transactions. 
Therefore, we explored ways to approximate the optimal ordering with reduced computational complexity.
We also want the ordering rule to be easy to verify, so that agents can monitor its implementation.  

We considered the \GSR method of  ~\cite{xavier2023credible}, which is described in Section~\ref{sec:Background}. 
The \GSR  has several significant advantages: it is simple to implement, computationally efficient, and verifiable.
Secondly, it is proven to be resilient against sandwich attacks. 
To test \GSR as a baseline, we employed the \VHGSR variant from \cite{ankile2023see}: it enforces a unique \GSR ordering per block by selecting at each step the smallest pending transaction adhering to the \GSR rule.
As demonstrated below in the experimental evaluation (Section~\ref{sec:simulation}), \GSR/\VHGSR may be far from optimal with respect to  price stability (minimizing volatility).
Therefore, we offer a new method, \clvr, that can perform better with a small computation cost. \footnote{
In this text we present a simple variation of the algorithm which incurs $O(n^2)$ computational cost.
This can be reduced to an $O(n\cdot \log\ n)$ computational cost, shown in Appendix~\ref{appendix:computational_cost}.
}

\subsubsection{CLVR Look-ahead Volatility-Minimizing (``\clvr\unskip'') Rule:}
\label{subsec:clvr}

Consider the following rule: at step $t$, select as the next trade the transaction that minimizes $(\ln~p_0~-~\ln~P(\sigma,~t))^2$.
That is, the rule picks at each step $t$ as next trade that causes minimal local one-step price-volatility from the status quo price, $p_0$.
In Appendix \ref{Sandwich Attacks} we provide a proof that \clvr protects traders from the traditional 3-transaction sandwich attacks.
Additionally, Section~\ref{subsec:Performance} shows that the \clvr is approximately optimal in many cases, even with larger numbers of pending transactions.
Appendix \ref{quality} provides an example and explains why CLVR is only approximately optimal and can deviate from optimality, but this deviation is negligible.



%% file: sections/05.Simulation.tex
\section{Simulations and Results}
\label{sec:simulation}

In this section we describe a series of 
experiments performed to show the effects of the economic measure (price stability) discussed in Section~\ref{sec:optimality}.\footnote{There are other economic measures that could be considered; as an example, we investigate inequality in Appendix~\ref{appendix:inequality}.}

\subsection{Optimal Ordering of Transactions}
\label{sec-Optimal ordering of transactions}
We first set a baseline database of synthetically generated blocks of transactions and compute the optimal orderings from a price-stability point of view. 

We aim to perform this experiment with synthetic trades that nonetheless closely mimic real-world workloads.
In order to achieve this, we collected all 4,192 swaps that executed on Uniswap V2's USDC-USDT AMM between June 19, 2024 and July 19, 2024 (UTC).\footnote{Smart contract: 0x3041CbD36888bECc7bbCBc0045E3B1f144466f5f}
We identify each swap's total quantity of tokens traded, and fit the data to a log-normal distribution, with parameters $\mu = 4.93$ and $\sigma = 2.05$.\footnote{See Figure~\ref{fig:trade_size_distribution} in Appendix~ \ref{appendix:Robustness checks}, which shows the goodness of fit of the swap distribution to a log-normal distribution.}
We also find that during this time period the AMM's mean reserves of USDC and USDT is about 2 million tokens, each.

We create a simulated AMM following the model as described in Section~\ref{sec:model}.
The AMM trades tokens X and Y, each starting with reserves $x_0 = y_0 = 2,000,000$ -- this also has the convenient effect of setting the initial spot price $p_0$ to 1.
For each trial in this experiment we generate a workload of $n$ pending transactions.
Each transaction is randomly assigned to either `sell' or `buy', each with 50\% probability.
The transaction's quantity of tokens is then drawn from the log-normal distribution described above.
In Appendix \ref{appendix:Robustness checks}, we run some robustness checks to ensure that our results can be generalized even when we have different amounts of liquidity available ($x_0,y_0$) or different distributions of the transaction sizes. 

\begin{figure}[htb]
  \centering
  \begin{minipage}{0.65\columnwidth}
    \includegraphics[width=\linewidth]{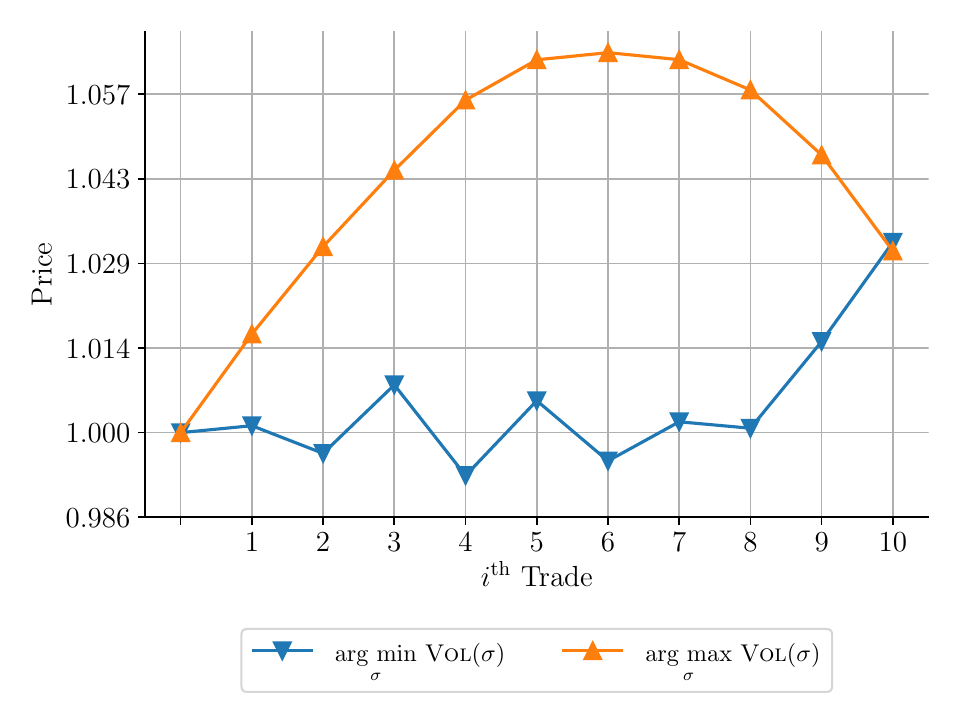}
  \end{minipage}\hfill
  \begin{minipage}{0.35\columnwidth}
    \caption{
      Price walk of different objective functions for a set $M$ of 10 synthetic transactions.
    Each point represents the AMM's spot price after executing the $i^{\textup{th}}$ transaction in the sequence.
    Furthermore, observe that for liquidity pools with sufficiently large liquidity (such as this), a block's closing price is \emph{not} sensitive to the chosen transaction ordering - leaving negligible impact on \emph{inter-}block volatility.
    }
    \label{fig:ex1}
  \end{minipage}
\end{figure}

When tractable, we use a $O(n!)$ brute-force search strategy to compute sequences that minimize volatility.
This experiment is written in approximately 500 lines of Rust code.
One such simulation is shown in Figure~\ref{fig:ex1}, which illustrates the dramatic difference in spot prices that occurs when minimizing (or maximizing) the price volatility.

\vspace{-15pt}

\subsection{A Comparison of \clvr with \VHGSR}
\label{subsec:Performance}

Our first set of experiments compares \VHGSR (the variant of \GSR we employ) with our proposed rule, \clvr, using as a benchmark the globally optimal and least-optimal ordering where computable.
This allows us to measure \emph{relative volatility}, which ranges from $100\%$ (global maximum) to $0\%$ (global minimum).
We ran simulations with varying numbers of transactions $n$ (block size), from 2 up to 1,000 transactions per block. 
The results are shown in Table~\ref{tab:Performance}, demonstrating that \clvr gives a better approximation to the optimal ordering for all tested block sizes (except the trivial case of $n \leq 2$).

The performance of \clvr in terms of \emph{relative volatility} suggests that \clvr approximately minimizes price volatility. Across simulations and different values of $n$, relative volatility values typically lie between $0\%$ and $0.2\%$, while the \emph{maximum} relative volatility across all simulations reaches at most $5.55\%$. In Section~\ref{subsec:worst}, we further provide a worst-case performance evaluation of \clvr and an estimate of an upper bound on its deviation from the optimal ordering.


\begin{table}[htb]
\centering
{\setlength{\tabcolsep}{5pt}
\begin{tabular}{@{}rrrrrrrrr@{}}
\toprule
 & \multicolumn{3}{c}{Best Price Stability (\%)} & \multicolumn{2}{c}{Mean Rel. Vol. (\%)} & \multicolumn{2}{c}{Max Rel. Vol. (\%)}  \\
\multicolumn{1}{r}{$n$} & \multicolumn{1}{r}{\clvr} & \multicolumn{1}{r}{\VHGSR} & \multicolumn{1}{r}{Tie} & \multicolumn{1}{r}{\clvr} & \multicolumn{1}{r}{\VHGSR} & \multicolumn{1}{r}{\clvr} & \multicolumn{1}{r}{\VHGSR} & \multicolumn{1}{r}{p-value} \\ \midrule
 2 & 0 & 0 & 100 & 0.00 & 0.00 & 0.00 & 0.00 & - \\
\rowcolor[HTML]{EFEFEF} 5 & 59 & 0 & 41 & 0.20 & 6.67 & 5.55 & 73.87 & 0.030 \\
 10 & 90 & 2 & 8 & 0.10 & 4.00 & 2.07 & 51.78 & 0.126 \\
\rowcolor[HTML]{EFEFEF} 12 & 96 & 0 & 4 & 0.08 & 3.96 & 1.19 & 67.94 & 0.023 \\
 50 & 98 & 2 & 0 & - & - & - & - & 0.091 \\
\rowcolor[HTML]{EFEFEF} 100 & 97 & 3 & 0 & - & - & - & - & 0.002 \\
 500 & 90 & 10 & 0 & - & - & - & - & 0.106 \\
\rowcolor[HTML]{EFEFEF} 1000 & 88 & 12 & 0 & - & - & - & - & $<0.001$ \\
\bottomrule
\end{tabular}
}
\captionof{table}{%
Comparison of \VHGSR vs \clvr. ``Relative volatility'' ranges from 100\% (global max) to 0\% (global min) -- computed when tractable. 100 trials performed for each block size $n$. The $p$-value represents a paired, one-sided t-test.%
}
\label{tab:Performance}
\end{table}

\begin{figure}[htb]
  \centering
  \begin{minipage}{0.6\columnwidth}
    \includegraphics[width=\columnwidth]{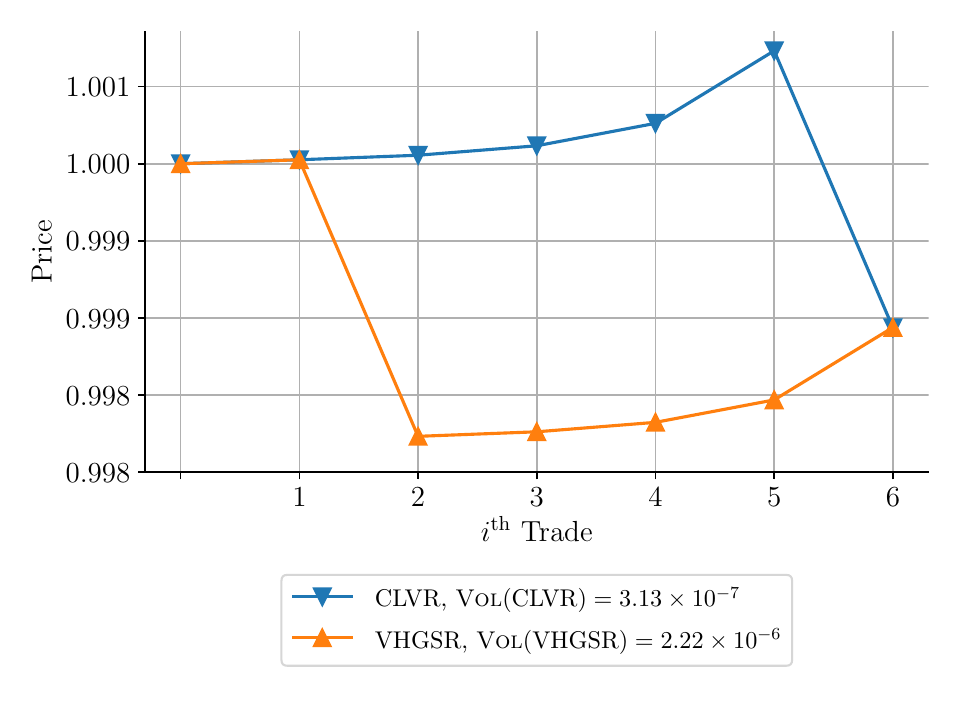}
  \end{minipage}\hfill
  \begin{minipage}{0.35\columnwidth}
    \caption{
      Example of a workload where \VHGSR performs poorly vs \clvr. \VHGSR is forced to place a large trade in position 2, which harms price stability.
    }
    \label{greedy}
  \end{minipage}
\end{figure}

In Figure~\ref{greedy} we visually demonstrate an example where, with respect to price volatility, there can be a significant difference between the \VHGSR and \clvr orderings.
In the example, the \VHGSR ordering deviates significantly from the baseline price as it is forced to execute a large, high-price-impact trade early in the sequence.


\subsection{Transaction failure rate}
\label{subsec:failure_rate}

In order to demonstrate the effectiveness of \clvr at reducing transaction failure rate due to users' slippage tolerance settings, we perform the following experiment.

First, we modify \clvr to be (na\"ively) slippage-tolerance aware.
Each trade is assigned a ``minimum amount out'', which must be received or else the transaction fails -- this represents the user's choice of slippage tolerance.
At each step, when \clvr determines the next suitable trade to sequence, the algorithm will disregard any trades which would fail if sequenced immediately.
If no viable trades are able to be sequenced then all remaining trades are sequenced as failed transactions.

Next, we assign to each synthetic trade a suitable slippage tolerance.
This is computed as follows: for each trade we simulate its execution in isolation against the ``status-quo'' price, and record the amount of tokens it receives for its fixed payment.
We reduce this amount by 0.5\% to achieve a ``minimum amount out.''
In practice, this reflects actual slippage tolerance setting behavior commonly observed on popular AMM applications~\cite{chemaya2023power}.
We compare \clvr and \VHGSR against a randomly generated set of trades and measure the percentage failure rate of each set in Table~\ref{tab:failure_rates}.
We notice that \clvr and \VHGSR are, on the whole, equivalent in performance with respect to minimizing transaction failures.


\begin{table}[ht]
\centering
\begin{minipage}{0.7\columnwidth}
{\setlength{\tabcolsep}{10pt}
\begin{tabular}{rrrrr}
\toprule
\multicolumn{1}{r}{} & \multicolumn{3}{c}{Failure Rate (\%)} & \multicolumn{1}{c}{\shortstack{\% \\Reduction\\ \space}} \\
$n$ & Random & VHGSR & CLVR & \shortstack{CLVR\\vs\\Random} \\
\midrule
3 & 1.73 & 0.07 & 0.07 & 96.2 \\ \rowcolor[HTML]{EFEFEF}
5 & 3.54 & 0.08 & 0.08 & 97.7 \\ 
8 & 6.31 & 0.26 & 0.26 & 95.8 \\ \rowcolor[HTML]{EFEFEF}
10 & 7.79 & 0.37 & 0.37 & 95.3 \\ 
15 & 11.85 & 0.53 & 0.53 & 95.6 \\ \rowcolor[HTML]{EFEFEF}
25 & 16.67 & 0.80 & 0.80 & 95.2 \\ 
50 & 22.92 & 0.97 & 0.98 & 95.7 \\ \rowcolor[HTML]{EFEFEF}
100 & 27.07 & 0.72 & 0.75 & 97.2 \\ 
\bottomrule
\end{tabular}
}
\end{minipage}\hfill
\begin{minipage}{0.28\columnwidth}
\captionof{table}{%
Failure rates of random and CLVR trade sequences for different numbers of trades, $n$. Trades were generated using the log-normal distribution discussed in Section~\ref{sec-Optimal ordering of transactions}. For each $n$ we run 1,000 trials and report the mean percentage failure rate.
}
\label{tab:failure_rates}
\end{minipage}
\end{table}

These results show a significant benefit for traders, with spurious transaction failures due to slippage protection reduced by about 95\%.
The reduction in failure rates has a significant economic impact on traders, as it not only increases the likelihood of executing a transaction and capturing the associated benefits but also helps avoid paying unnecessary gas fees. Table~\ref{tab:failure_rates} further shows that if AMMs scale up and accommodate more transactions per block, failure rates become a significant concern. In our simulations, with random ordering, the failure rate increases from 7.8\% to 27\% as the number of transactions per block rises from 10 to 100. Additionally, the benefits extend beyond traders, in a real-world setting with trading fees this also increases payments made to liquidity providers, as no fee is collected if the transaction fails.



\subsection{Price Stability and Block Sizes}
\label{sub:Blocksize}

The next set of experiments examines the impact of block size on price stability under \clvr ordering.
In traditional financial markets, there is an extensive discussion on what the optimal trading frequency is \cite{du2017optimal,budish2014implementation,jagannathan2022frequent,fricke2018too}.
Trading frequency means setting the time that the market will operate.
When performing batch auctions, for example, it sets the duration of the auction.
Setting the trading frequency can directly impact the number of transactions in a given batch auction.
High-speed frequency can be seen as a first come, first serve system, while a slower frequency can increase the number of transactions in the batch.
In that sense, we can think about ``block size'' as the duration that the social planner sets for trading frequency on an AMM.
Having a lower frequency can allow more trades to be executed in a given block, which we say are ``bigger'' blocks. 

Our evaluation strategy is as follows. 
We synthesize a set of 100 transactions using the process described above (in section \ref{sec-Optimal ordering of transactions}). 
Each transaction is given a sequential timestamp, which indicates the order of the transaction's arrival in the ``mempool''.
For simplicity, we will assume that the spot market price is equal to 1, as we set in the previous simulations.

Now we can test what will happen if the social planner executes all the transactions using a first come, first served policy, which can be seen as having 100 blocks (one transaction per block).
We also examine what would happen if there were batches of different block sizes, for example, ten blocks with ten transaction each, or one big block with 100 transactions.
We test this over a variety of block sizes.

\begin{figure}[htb]
  \centering
  \begin{minipage}{0.6\columnwidth}
    \includegraphics[width=\columnwidth]{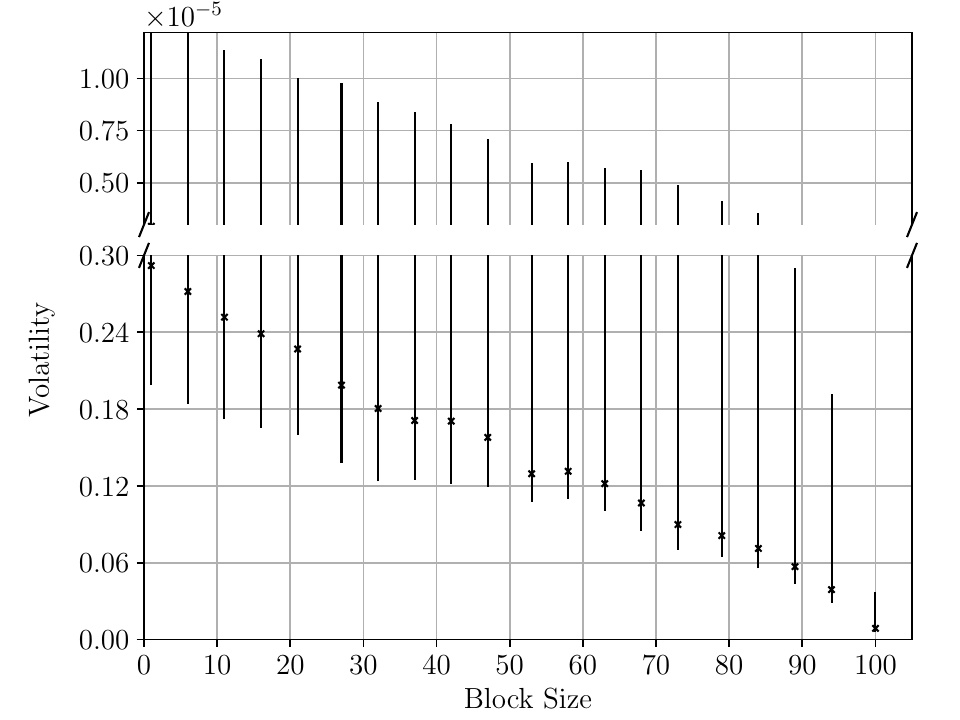}
  \end{minipage}\hfill
  \begin{minipage}{0.35\columnwidth}
    \caption{
      Price volatility achieved by \clvr with varying block sizes. Computed over 1,000 trials, error bars drawn at 25th and 75th percentiles, marker placed at median.
    }
    \label{Blocksize}
  \end{minipage}
\end{figure}

Figure~\ref{Blocksize} shows the results of running this simulation over 1,000 trials, demonstrating the advantage in price stability with batching transactions in larger blocks.
When we have a first come, first serve approach (on the left), this gives the highest median price volatility with a very high variance.
In contrast, price volatility is significantly reduced when all transactions are batched into one big block (on the right).
This allows the \clvr ordering mechanism to achieve a much lower median of price volatility with much lower variance.
In summary, transaction ordering has a significant impact on price volatility when the block size is very large.
A social planner may want to take this into consideration when deciding the trading frequency.
\vspace{-10pt}

\subsection{Splitting Transactions}
\label{sec:Splitting}
Traders can benefit from splitting their transactions into smaller ones, also known as order splitting \cite{park2023conceptual}. 
While this increases gas costs (as we explain further in Section \ref{sec:limitations}), it also improves price stability and execution outcomes under our mechanism. 
For example, splitting a buy and a sell of 1{,}000 units each into 1{,}000 smaller orders and interleaving them reduces volatility by six orders of magnitude.

We conduct simulations to quantify the benefit of splitting: both individually (a single trader splits) and collectively (all traders split). 
In both cases, we find consistent improvements in execution quality. Full experimental setup, figures, and quantitative results are provided in Appendix~\ref{appendix:splitting}.

\subsection{\clvr Applied to Empirical Data}
\label{Empirical-Data}

\begin{table}[htb]
\centering
\caption{
Comparison of \VHGSR vs \clvr over real-world blocks.
The $p$-value is computed from a paired, one-sided t-test.
}
\label{tab:volatility-empirical}
\begin{tabular}{@{}ccccccccc@{}}
\toprule
\multicolumn{2}{c}{} & \multicolumn{3}{c}{Winner Count} & \multicolumn{3}{c}{Mean Relative Volatility} & \\
Swap Count & Blocks & \VHGSR & \clvr & Tie & Current & \VHGSR & \clvr & $p$-value \\ \midrule
3 & 1432 & 0 & 263 & 1169 & 55.45 & 7.74 & 0.16 & $<0.001$ \\
\rowcolor[HTML]{EFEFEF} 4 & 277 & 0 & 95 & 182 & 52.84 & 9.05 & 0.24 & 0.092 \\
5 & 66 & 1 & 25 & 40 & 44.11 & 6.84 & 0.14 & 0.105 \\
\bottomrule
\end{tabular}
\end{table}

We evaluate \clvr's performance on the 4,192 swaps collected above (Uniswap V2’s USDC-USDT AMM).
We start with an AMM that has reserves $x_0 = y_0 = 2,000,000$.
Then, we compute the volatility after executing each transaction in its observed, real-world ordering.
Since this pool allows traders to trade two stablecoins, their exchange rate is relatively stable $(p_0=1)$, which theory suggests impacts trading frequency~\cite{du2017optimal}.
Indeed, we observe that most blocks with trading activity contain only a single trade, in which case transaction ordering is irrelevant.
To simulate increased AMM trading activity, we collect the swaps into 420 blocks of size 10, ordered in their real-world ordering, and execute \clvr in a multi-round fashion.
We measure an 85\% reduction in volatility, which shows the potential benefit of implementing ordering mechanisms on AMMs.

We also evaluate \VHGSR vs. \clvr when applied to the Uniswap V2 USDC-WETH AMM swaps, collected over the same time range as above.\footnote{Smart contract: 0xB4e16d0168e52d35CaCD2c6185b44281Ec28C9Dc}
This pool has a much higher trading frequency compered to Uniswap V2’s USDC-USDT AMM, which allows us to test \VHGSR vs. \clvr performances on blocks with more than 3 swaps. 
We compute block-level performance statistics and aggregate by the block's swap count.
Blocks containing more than 5 swaps are exceedingly rare (fewer than 30 samples per swap count), and are thus omitted to avoid spurious outcomes.
This is evaluated in Table~\ref{tab:volatility-empirical}, where we also find that \clvr achieves good performance, with significant price volatility reduction compared to existing volatility.\footnote{The p-values for 4–5 swaps are lower mainly due to the reduction in the number of blocks, which substantially decreases from 1432
for 3 swaps to 277 (66)
for 4 (5) swaps.}





\subsection{CLVR Worst-Case Performance and Upper-Bound Estimation}
\label{subsec:worst}

While our first set of experiments in Section~\ref{sec:simulation} suggests that \clvr achieves very low levels of price volatility compared to the optimal ordering, it is still important to discuss the potential upper bounds of this approximation and to characterize the possible gap from optimality associated with using \clvr .

Specifically, across multiple experiments, including settings in which we change the AMM trading environment by varying the available liquidity or altering the transaction size distribution of traders (see Appendix~\ref{appendix:Robustness checks}, Robustness Checks)---we compare \clvr ’s price deviation to the optimal ordering for different numbers of transactions to be ordered. We find consistent results across all experiments and transaction counts: on average, the mean relative volatility lies between $0\%$ and $0.67\%$ (see Tables~ \ref{tab:Performance},~
\ref{tab:volatility-empirical}, and~\ref{tab:robustness}). Our goal is to identify which settings can affect the optimality gap of the \clvr  approximation.

To address this question, we first identify a set of cases in which \clvr  yields the optimal ordering and minimizes price deviation. We then conduct a worst-case \footnote{‘worst-case’ refers to the maximum observed deviation across all simulated transaction sets under a fixed environment.} performance analysis to estimate upper bounds on the gap induced by using \clvr  as an approximation. Finally, we examine how pool size (i.e., available liquidity) and transaction size affect these upper-bound estimates.

From Table~\ref{tab:Performance}, it is clear that the number of transactions awaiting execution can
significantly impact the quality of the approximation. In particular, under certain
conditions, \clvr  coincides with the optimal ordering and fully minimizes price
volatility. For example, when the number of transactions to be ordered is sufficiently small ($n \leq 2$),
\clvr  always produces the optimal ordering.

These results are intuitive. When $n=1$, ordering is irrelevant, and all ordering
algorithms yield the same outcome. When $n=2$, the optimal ordering that minimizes
price volatility executes the smaller transaction first, followed by the larger one.
Both \clvr  and VHGSR follow this rule, and therefore coincide with the optimal ordering.
In addition, when $N>2$ and all transactions move prices in the same direction, the
optimal strategy is to execute transactions from smallest to largest in order to
minimize the average price deviation across all transactions. This result follows from the fact that, in an AMM with a constant-product market maker (CPMM) price function, the marginal execution price is increasing.

For all remaining cases in which \clvr deviates from the optimal ordering, we aim to
estimate upper bounds on these deviations. The magnitude of the deviation depends on
the specific set of pending transactions, including the number of transactions, their
sizes, and their directions, as well as on the amount of liquidity available in the
AMM. Repeated simulations with different sets of pending transactions allow us to
explore such cases and provide empirical estimates of these upper bounds.

To better estimate these upper bounds, Table~\ref{tab:Performance} indicates that CLVR’s maximum observed relative volatility occurs when the number of pending transactions is $n = 5$, with a maximum relative volatility of approximately $5.5\%$. These results suggest that the largest deviation occurs when the number of transactions to be executed is around five, a pattern that is also consistent with other simulation results under different liquidity levels and transaction distributions reported in Tables~\ref{tab:volatility-empirical}, and~\ref{tab:robustness}.

The intuition is as follows. When the number of pending transactions is sufficiently small (e.g., $n \leq 2$), \clvr minimizes price volatility, as discussed earlier. As the number of pending transactions increases beyond two, the \clvr ordering algorithm begins to deviate from optimality. However, when the number of transactions becomes sufficiently large (e.g., $n \geq 10$), the algorithm becomes more flexible in rearranging transactions to reduce price volatility, as discussed in Subsections ~\ref{sub:Blocksize} and ~\ref{sec:Splitting}. As a result, when many pending transactions must be ordered, relative volatility values under the \clvr algorithm tend to be closer to the optimal ordering rather than to the maximum volatility, implying that \clvr performs well under the relative volatility metric.

\begin{figure}[htb]
  \centering
  \begin{minipage}{0.6\columnwidth}
    \includegraphics[width=\columnwidth]{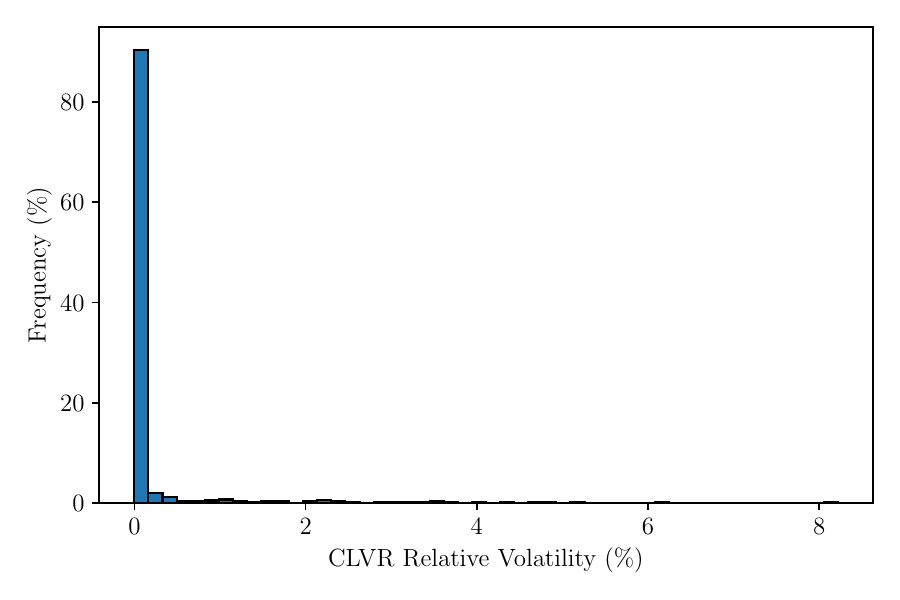}
  \end{minipage}\hfill
  \begin{minipage}{0.35\columnwidth}
    \caption{
      Distribution of \clvr's relative volatility, which ranges from $100\%$ (global maximum) to $0\%$ (global minimum), across $1{,}000$ simulations with $n = 5$ pending transactions.
    }
    \label{clvr_relative_volatility_histogram_5_trades}
  \end{minipage}
\end{figure}

To further explore and estimate \clvr’s gap from optimality, we run $1{,}000$ simulations with $n = 5$, where the largest observed deviation occurs at this intermediate block size.
We then plot the distribution of \clvr’s relative price volatility.
Figure~\ref{clvr_relative_volatility_histogram_5_trades} illustrates that relative volatility is close to zero in most cases, while the maximum observed value reaches approximately $8\%$.
Finally, given the CPMM pricing rule, we know that another important factor influencing price volatility is the transaction size relative to the available liquidity in the pool, as explained in \cite{chemaya2022estimating}. When the ratio of transaction size to pool liquidity increases, the CPMM pricing rule yields a higher execution price, and this ratio can therefore affect the optimality of \clvr when it varies.

\begin{figure}[htb]
  \centering
  \begin{minipage}{0.6\columnwidth}
    \includegraphics[width=\columnwidth]{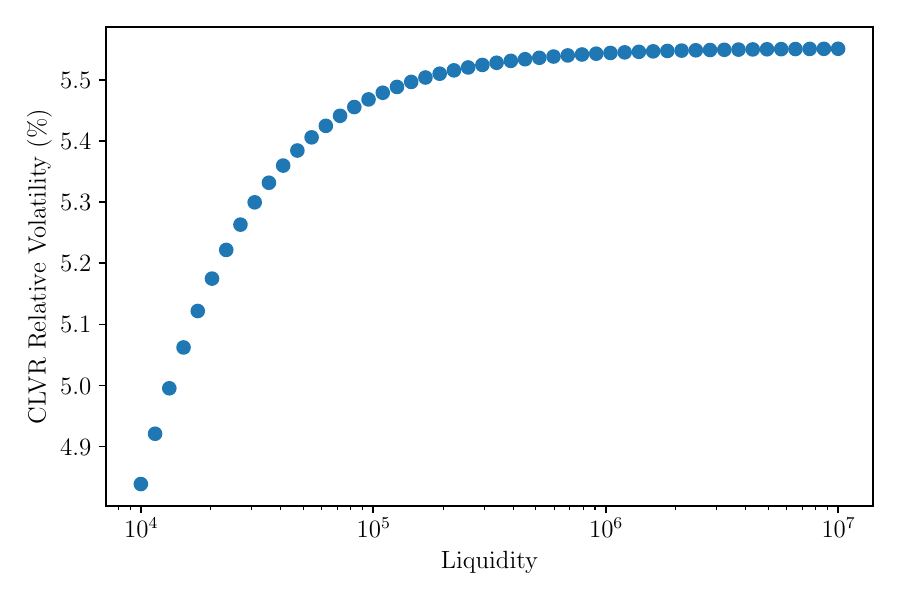}
  \end{minipage}\hfill
  \begin{minipage}{0.35\columnwidth}
    \caption{The y-axis represents CLVR’s relative volatility for the worst-case set of pending transactions identified in Table~\ref{tab:Performance}, while the x-axis shows how price volatility changes as the amount of liquidity in the pool varies for this specific set. Liquidity is measured as $x_0 = y_0$, the reserves of the AMM.}
    \label{figures/clvr_relative_volatility_vs_liquidity}
  \end{minipage}
\end{figure}

To explore this dimension, we select several specific sets of pending transactions in which \clvr exhibits relatively high price volatility (i.e., poor performance) and examine how relative price volatility changes as we vary the liquidity available in the pool. Figure~\ref{figures/clvr_relative_volatility_vs_liquidity} presents the results for a representative example. The figure suggests that while reducing liquidity generally increases price volatility, it can also improve CLVR’s relative performance: the \clvr algorithm tends to produce outcomes closer to the optimal ordering rather than to the maximum volatility, implying that \clvr performs well under the relative volatility metric. In contrast, increasing the liquidity in the pool appears to have only a marginal impact on CLVR’s relative volatility.

%% file: sections/06.Conclusions.tex
\section{Conclusions}
\label{sec:conclusion}
In this work we demonstrate that transaction ordering can be treated as a social good, serving to not just prevent harm (from sandwich attacks, etc) but also to benefit traders.
We focus on the economic goal of maintaining intra-block price stability:
transaction ordering that minimizes intra-block price volatility benefits traders by offering more stable prices and reducing transaction failure rates.
To this end, we propose the \clvr algorithm, which efficiently computes an approximately volatility-minimizing sequence and exhibits very low deviation from optimality.
We show that \clvr outperforms prior work (\VHGSR) by reducing price volatility.
\clvr also significantly improves the transaction failure rate compared to random transaction ordering.
In addition, we show that a social planner using \clvr may want to order transactions in bigger blocks (a larger number of transactions) to improve the system's price stability.
Finally, we show that traders benefit from splitting their transactions under \clvr.

This work has broad implications for AMM design and, in general, trading market design.
In the DeFi space, AMM protocol designers can adopt our ordering approach by enforcing \clvr ordering via smart contracts such as a Uniswap V4 hook\cite{adams2023uniswap,bachu2025overview}.
At the blockchain level, mechanisms like Protected Order Flow (PROF) can commit to ordering rules like CLVR and enforce its implementation~\cite{babel2024prof}.
Beyond the DeFi space, our social planner could also manage an AMM run by a centralized authority.
Private AMMs are one possible direction when adopting AMMs to traditional markets.
In that case, the AMM can be regulated, and an ordering mechanism will be enacted and enforced.

Future research may explore the impact of transaction ordering in AMMs, which directly affects traders. As the number of transactions increases, so does the importance of ordering. If AMMs scale up, transaction ordering must be taken seriously, as ordering mechanisms have a strong potential to improve AMM performance.

\subsection{Limitations and Future Directions}
\label{sec:limitations}

\newpar{Practical trading aspects}
We do not account for traders using routers, which split transactions over several liquidity pools -- this would require a multi-pool analysis.

\newpar{Protocol complexities}
If the social planner is able to censor or insert transactions, there may be implications on iterated (multi-block) sequencing which circumvent \clvr's ordering capabilities.
There is promising concurrent work on anti-censorship measures, such as inclusion lists, which tackle this problem \cite{EIP-7547,ether,babel2024prof}.

\newpar{Gas fees}
We also do not account for a protocol's gas prices and fees when considering a trader's decision to split their transaction.
Transaction splitting can also be viewed as equivalent to increasing block size, in the sense that once transactions are split into many smaller pieces, the effect is similar to increasing the number of transactions per block.
Future work may examine the impact of fees on traders' strategic decision-making and also the impact on the block size.

\newpar{Alternative economic goals}
Price stability is not the only economic goal that a social planner may want to achieve.
In Appendix~\ref{appendix:inequality} we examine an alternate goal: reducing inequality.
In a nutshell, a social planner that is minimizing inequality seeks to give ``good'' prices to the less-wealthy traders at the expense of wealthy traders.
We show that this goal is incompatible with maintaining price stability -- they cannot be simultaneously pursued.
Further frustrating this objective is the fact that inequality is a somewhat illusive concept on blockchains, since wallet ownership is anonymized and multiple wallets can be under the control of a single user (a \emph{sybil} attack).
It is left for future research to further investigate the impact of transaction orderings on inequality, coupled with the challenge of analyzing wealth ownership on blockchains.  

\newpar{Future directions}
Transaction ordering may have an impact on a market's price discovery over time, especially if a social planner decides to increase the block-time.
Enforcing an ordering mechanism may also have implications for the activities of informed traders (e.g., centralized exchange arbitrageurs).
Future work should analyze the impact of \clvr on liquidity providers and how it mitigates Loss-Versus-Rebalancing (LVR) \cite{milionis2022automated}.

%% file: sections/XX.Appendix.tex
\section{Other Economics measures}
\label{appendix:inequality}

\subsection{Inequality}
\label{subsec:Inequality}

Other economics measurement could be of interests instead of price volatility. In this subsection, we consider inequality. We consider ordering strategy that prioritizes certain traders by organizing trades so that favored participants receive higher prices when selling and lower prices when buying.

Inequality can be evaluated in different ways, depending on the assumptions and objectives of the policymaker. 
One approach assumes that transaction sizes correlate with users' initial wealth, and orders trades to favor smaller transactions.


Another approach evaluates inequality based on total user wealth, measured through wallet balances. 
However, this is challenging in the pseudonymous blockchain environment, where users can control multiple accounts to conceal their true wealth (i.e., a \emph{sybil} attack).



\subsubsection{Traders' Inequality Based On Their Transaction Size - Gini Post Trade Wealth:}
For a given set of pending transactions $M$ in a given block, the transaction order that reduces traders' inequality while accounting for transaction size is the one that results in the lowest Gini coefficient (lowest inequality) after the transactions are executed.\footnote{The Gini Coefficient is a well-known economic measurement of inequality. Gini Coefficient ranges from 0 to 1, with higher values indicating greater inequality \cite{yitzhaki2013gini}.} Intuitively, this works by ensuring 
that smaller transactions receive a ``good" exchange price, while larger transactions receive a ``bad" exchange price, thus reducing inequality from both ends.

In a simple example when the exchange rate between token X and token Y is one-to-one ($p_0=1$), the amount of tokens each trader redeems from a transaction ($x_{\textup{out}}(\sigma,i)$ or $y_{\textup{out}}(\sigma,i)$) can represent its ``wealth". The ``wealth" and corresponding Gini Coeffient are given in the formulas below:
\begin{align}
W_i &= x_{\textup{out}}(\sigma,~j) \ or \ y_{\textup{out}}(\sigma,~j) \\
 \textsc{Gini}(\sigma) &= \frac{2 \cdot \sum_{i=1}^{n} i \cdot W_i}{n \cdot \sum_{i=1}^{n} W_{i}} - \frac{n + 1}{n}
\end{align}

Where $W_{1}\leq W_{2}\leq \cdots \leq W_{n}$.
The social planner will need to minimize the Gini coefficient by ordering the transactions in $M$ in a way that minimizes this function.

Finally, Table~\ref{tab:summary-statistics-comparison} shows summary statistics of both price stability and equality objectives from all simulations, which demonstrates that optimizing for one objective produces poor outcomes for the other objective.

{\setlength{\tabcolsep}{5pt}
\begin{table}[H]
\centering
\caption{
Relative optimization scores of price stability and equality objectives.
Each score is computed as a relative percentage, where 0\% is the global minimum, and 100\% is the maximum.
Taken as the mean of 100 trials.
We show the relative Gini coefficient when optimizing for price stability, and likewise the relative volatility when optimizing for equality.
}
\begin{tabular}{@{}llllll@{}}
\toprule
 & \multicolumn{5}{c}{Number of Trades} \\
 & \multicolumn{1}{r}{3} & \multicolumn{1}{r}{5} & \multicolumn{1}{r}{10} & \multicolumn{1}{r}{12} & \multicolumn{1}{r}{13} \\ \midrule
 & \multicolumn{5}{c}{Relative Gini Coefficient} \\
Maximizing Price Stability & 45.02 & 45.30 & 48.76 & 46.93 & 48.89 \\
Minimizing Price Stability & 56.47 & 55.08 & 52.25 & 53.05 & 51.88 \\ \midrule
 & \multicolumn{5}{c}{Relative volatility} \\
Maximizing Equality & 48.53 & 50.97 & 52.64 & 50.71 & 50.31 \\
Minimizing Equality & 47.31 & 43.07 & 37.93 & 37.98 & 38.05 \\
\bottomrule
\end{tabular}
\label{tab:summary-statistics-comparison}
\end{table}}

\clearpage

\section{Model Equations}
\label{appendix:model_equations}

The following equations compute the amount received when the $i^{\textup{th}}$ trade in ordering $\sigma$ is executed on a constant-product automated market maker with initial reserves $x_0$ and $y_0$.

\begin{align}
    y_\textup{out}(\sigma, i) = \begin{cases}
        x_\textup{in} \frac{Y(\sigma,~i-1)}{X(\sigma,~i-1) + x_\textup{in}} &\textup{if } \sigma(i) = \textup{sell }x_\textup{in} \\
        0 &\textup{otherwise}
    \end{cases} \\
    x_\textup{out}(\sigma, i) = \begin{cases}
        y_\textup{in} \frac{X(\sigma,~i-1)}{Y(\sigma,~i-1) + y_\textup{in}} &\textup{if } \sigma(i) = \textup{buy }y_\textup{in} \\
        0 &\textup{otherwise}
    \end{cases}
\end{align}

Where functions $X(\sigma,~i)$ and $Y(\sigma,~i)$ refer to the reserves of tokens X and Y after trade $i$, as follows:

\begin{align}
X(\sigma, i) &= \begin{cases}
  x_0 & \textup{if } i = 0 \\
  X(\sigma,~i-1) + x_\textup{in} & \textup{if } i > 0 \textup{ and } \sigma(i) = \textup{ sell } x_\textup{in} \\
  X(\sigma,~i-1) -  x_{\textup{out}}(\sigma, i) & \textup{if } i > 0 \textup{ and } \sigma(i) = \textup{ buy } y_\textup{in} \\
\end{cases} \\
Y(\sigma, i) &= \begin{cases}
  y_0 & \textup{if } i = 0 \\
  Y(\sigma,~i-1) + y_\textup{in} & \textup{if } i > 0 \textup{ and } \sigma(i) = \textup{ buy } y_\textup{in} \\
  Y(\sigma,~i-1) -  y_{\textup{out}}(\sigma, i) & \textup{if } i > 0 \textup{ and } \sigma(i) = \textup{ sell } x_\textup{in} \\
\end{cases}
\end{align}





\clearpage

\section{Sandwich Attacks}
\label{Sandwich Attacks}
The \clvr algorithm prevents 3-transaction sandwich attacks.
In this section we provide a proof.

The three transactions involved in a sandwich attack are (1) the \emph{front-running} transaction, (2) the \emph{victim} transaction, and (3) the \emph{back-running} transaction.
The front-running and back-running transactions are inserted by an attacker in order to manipulate the victim into buying at an inflated price.
The attacker is thus able to extract risk-free profit.

In an attack, the transactions execute in exactly the order listed above.
In order for this manipulation to be profitable, the front-running transaction must push the price toward one less favorable for the victim, thus causing them to overpay.
Finally, the back-running transaction swaps in the opposite direction in order to recover a risk-free profit denominated in the token originally sold to the AMM.

Without loss of generality (see Section~\ref{subsec:Price Stability}), consider that the victim transaction sells token X to the AMM in order to receive token Y.
Then, the front-running transaction also sells X to receive Y, and the back-running transaction sells Y to receive X.

We define the following variables.
The amounts paid to the AMM by the front-running, and back-running transactions are $\Delta x_f$ and $\Delta y_b$, respectively, and the amounts received from the AMM are $\Delta y_f$  and $\Delta x_b$, respectively (the victim transaction size is not important for the proof).
Let $x_0$ and $y_0$ be the liquidity pool's reserves of token X and token Y, respectively, before transactions execute.
Let $p_0 = y_0 / x_0$ be the initial price.

The attacker receives a risk-free profit when $\Delta x_f < \Delta x_b$ and $\Delta y_b \leq \Delta y_f$ -- i.e., when the front-running transaction pays less X than the back-running transaction receives, \emph{and} the attacker's back-running transaction does not spend more token Y than the attacker received when executing the front-running transaction.
The former condition guarantees profit, and the latter condition guarantees that it is risk-free -- i.e., the attacker does not lose any token Y.

Now, our task is determine whether it is possible to perform a risk-free sandwich attack when also subject to the \clvr sequencing rule.
We proceed by contradiction.
Start by considering how to sequence the three transactions involved in the sandwich attack.
According to the sequencing rule, the first transaction must move the log-price the least.

Let $p_f$ and $p_b$ be the prices after executing the front-running and back-running transactions first (before any other transactions execute), respectively.
scThen we must ensure that $\left( \ln p_0 - \ln p_f \right)^2 \leq \left( \ln p_0 - \ln p_b \right)^2$ (or else the back-running transaction would be sequenced first, violating the sandwich attack order).
Since we know that $p_f < p_0 < p_b$ (given the trade directions), it must be that $\ln p_0 - \ln p_f \leq \ln p_b - \ln p_0$ or, equivalently, $p_0^2 \leq p_b p_f$, which is expressed as:

\begin{align}
\left(\frac{y_0}{x_0}\right)^2 &\leq \left(\frac{\Delta y_b + y_0}{- \frac{\Delta y_b x_0}{\Delta y_b + y_0} + x_0}\right) \left(\frac{- \frac{\Delta x_f y_0}{\Delta x_f + x_0} + y_0}{\Delta x_f + x_0}\right) = \frac{\left(\Delta y_b + y_0\right)^{2}}{\left(\Delta x_f + x_0\right)^{2}} \\
\end{align}

Since all values are positive,

\begin{align}
\label{eq:simple_price_ineq}
\frac{y_0}{x_0} &\leq \frac{\Delta y_b + y_0}{\Delta x_f + x_0}
\end{align}



Then, we must also satisfy $\Delta y_b \leq \Delta y_f$ (the back-running transaction cannot spend more Y than the front-running transaction purchases, or else violate risk-free profit).
Expanding $\Delta y_f$ yields:

\begin{align}
\label{eq:ineq_expanded_risk_free}
\Delta y_b \leq \frac{\Delta x_f y_0}{\Delta x_f + x_0}
\end{align}

Combining Inequality~\ref{eq:simple_price_ineq} and Inequality~\ref{eq:ineq_expanded_risk_free} gives:

\begin{align}
\frac{y_0}{x_0} &\leq \frac{\frac{\Delta x_f y_0}{\Delta x_f + x_0} + y_0}{\Delta x_f + x_0} \\
\frac{y_0}{x_0} &\leq \frac{y_0 \left(2 \Delta x_f + x_0\right)}{\left(\Delta x_f + x_0\right)^{2}} \\
\left(\Delta x_f + x_0\right)^{2} &\leq x_0 \left(2 \Delta x_f + x_0\right) \\
\Delta x_f^{2} + 2 \Delta x_f x_0 + x_0^{2} &\leq 2 \Delta x_f x_0 + x_0^{2} \\
\Delta x_f^2 &\leq 0
\end{align}

which is a contradiction, as $\Delta x_f$ is a positive real value.

\clearpage

\section{On the Quality of \clvr:}
\label{quality}

\begin{figure}[!ht]
    \centering
    \includegraphics[width=0.48\linewidth]{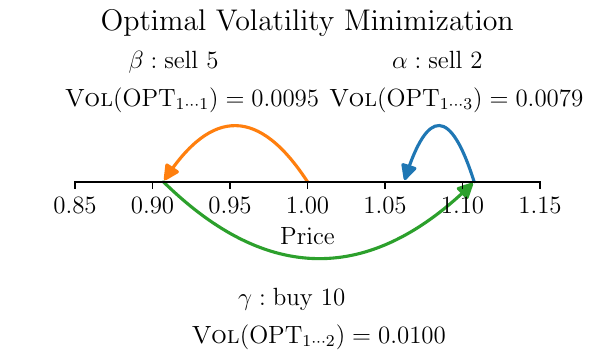} \hfill
    \includegraphics[width=0.48\linewidth]{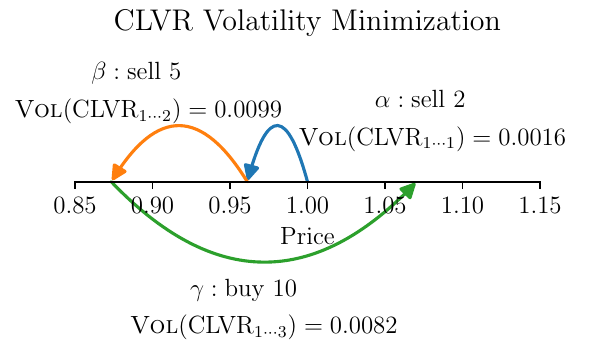}
    \caption{Optimal sequence minimizing price volatility, OPT (left) vs \clvr sequence (right)}
    \label{fig:LPmin-nonoptimal}
\end{figure}

A natural question to ask is whether, given a block of transactions, the \clvr rule always achieves the globally minimal price volatility.
Unfortunately, we can find a counter examples showing it does \textbf{not}.
For example, consider three trades: 
$\{\alpha:\textup{ sell }2,~\beta:\textup{ sell }5,~\gamma:\textup{ buy }10\}$
when the initial liquidity available is $x_0=y_0=100$.
The optimal ordering for minimizing aggregate price-volatility is \textbf{not} to start with sell 2, as \clvr would dictate.
Rather, it is sell 5, buy 10, sell 2.
The \clvr ordering is illustrated in Figure~\ref{fig:LPmin-nonoptimal} (right) against the optimal volatility-minizing ordering (left).
In this case, the average price volatility of the \clvr is $8.2\times 10^{-3}$, while the optimal ordering can reduce this volatility to $7.9\times 10^{-3}$.

This example highlights why \clvr only approximately minimizes price volatility.
However, even in this example, we can still see that the \clvr order gives very low price volatility, which is nonetheless close to optimal.
In section ~\ref{subsec:Performance}, we provide evidence that \clvr achieves nearly-optimal orderings in many cases. 

\clearpage

\section{Robustness checks}
\label{appendix:Robustness checks}
In this section, we run some robustness checks. 
We (1) change the liquidity available in the pool, and (2) change the size distribution from log normal to uniform.

\begin{table}[!htb]
\centering
\caption{Comparison of \VHGSR vs \clvr. When tractable, relative volatility is computed as a percentage where 0\% is the minimum volatility, and 100\% is the maximum.
The $p$-value is computed from a paired, one-sided t-test.
}
\label{tab:robustness}
\begin{tabular}{@{}cccccccc@{}}
\toprule
\multicolumn{7}{l}{Lower Liquidity $x_0 = y_0 = 100,000$} \\ \midrule
\multicolumn{1}{c}{} & \multicolumn{3}{c}{Winner Count} & \multicolumn{2}{c}{Mean Relative Volatility} \\
Block Size & VHGSR & \clvr & Tie & \VHGSR & \clvr & $p$-value \\ \midrule 
 3 & 0 & 18 & 82 & 8.55 & 0.27 & 0.078 \\
\rowcolor[HTML]{EFEFEF}  5 & 0 & 58 & 42 & 10.14 & 0.19 & 0.017 \\
 10 & 3 & 89 & 8 & 2.96 & 0.08 & 0.009 \\
\rowcolor[HTML]{EFEFEF}  12 & 3 & 93 & 4 & 1.74 & 0.07 & 0.021 \\
 13 & 4 & 88 & 8 & 2.36 & 0.09 & 0.038 \\ \midrule
\multicolumn{7}{l}{Uniform Distribution, trade size selected uniformly on $(0,~100,000)$} \\ \midrule
\multicolumn{1}{c}{} & \multicolumn{3}{c}{Winner Count} & \multicolumn{2}{c}{Mean Relative Volatility} \\
Block Size & VHGSR & \clvr & Tie & \VHGSR & \clvr & $p$-value \\ \midrule 
 3 & 0 & 9 & 91 & 3.12 & 0.20 & 0.004 \\
\rowcolor[HTML]{EFEFEF}  5 & 1 & 37 & 62 & 2.78 & 0.67 & $<0.001$ \\
 10 & 6 & 61 & 33 & 0.75 & 0.40 & $<0.001$ \\
\rowcolor[HTML]{EFEFEF}  12 & 8 & 76 & 16 & 0.60 & 0.30 & $<0.001$ \\
 13 & 11 & 73 & 16 & 0.43 & 0.27 & $<0.001$ \\
\bottomrule
\end{tabular}
\end{table}

\begin{figure}[htb]
  \centering
  \begin{minipage}{0.6\columnwidth}
    \includegraphics[width=\columnwidth]{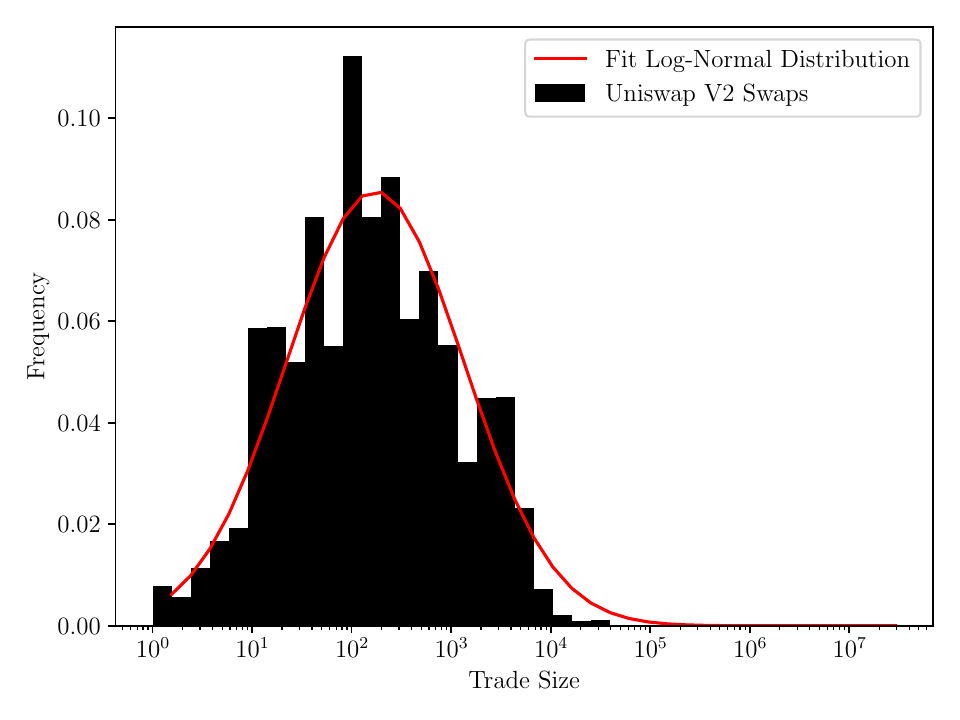}
  \end{minipage}\hfill
  \begin{minipage}{0.35\columnwidth}
    \caption{
      Distribution of 4,192 swaps executed on Uniswap~V2’s USDC--USDT AMM between June~19,~2024 and July~19,~2024 (UTC) . The figure also illustrates the goodness of fit of the swap size distribution to a log normal distribution with parameters $\mu = 4.93$ and $\sigma = 2.05$.
}
    \label{fig:trade_size_distribution}
  \end{minipage}
\end{figure}

\newpage

\section{Splitting}
\label{appendix:splitting}

Traders may split their transactions into small trades, also known as order splitting~\cite{park2023conceptual}, to strategically improve their exchange prices and avoid price manipulations.
In addition, splitting transactions into smaller ones provides more flexibility with respect to ordering possibilities, which helps improve the system's price stability. However, transaction splitting also entails costs, since traders must pay gas fees for each transaction. As a result, splitting incurs significantly higher gas costs than executing a single transaction, as we explain further in Section~\ref{sec:limitations}. Here, we focus on the benefits of transaction splitting for execution and price stability.

We can explain this idea with the following example: suppose there are only two pending transactions in a block.
One is buying ($\alpha:~\textup{buy}~1,000$), and the other is selling ($\beta:~\textup{sell}~1,000$).
Suppose that $x_0=y_0=100,000$ and, thus, $p_0=1$.
In this example, if we execute both transitions, we will have a price volatility of $2.0\times 10^{-4}$.
However, if both traders split their transactions into 1000 small transactions of 1 each, an optimizing algorithm can improve the price volatility to $2.0\times 10^{-10}$ -- an improvement of 6 orders of magnitude.
The optimal ordering interleaves smaller buy and sell orders, alternating between each, in order to maintain the status-quo price.
The more users split their transactions, the more this price volatility will decrease and eventually converge to 0.

\begin{figure}[htb]
  \centering
  \begin{minipage}{0.6\columnwidth}
    \includegraphics[width=\columnwidth]{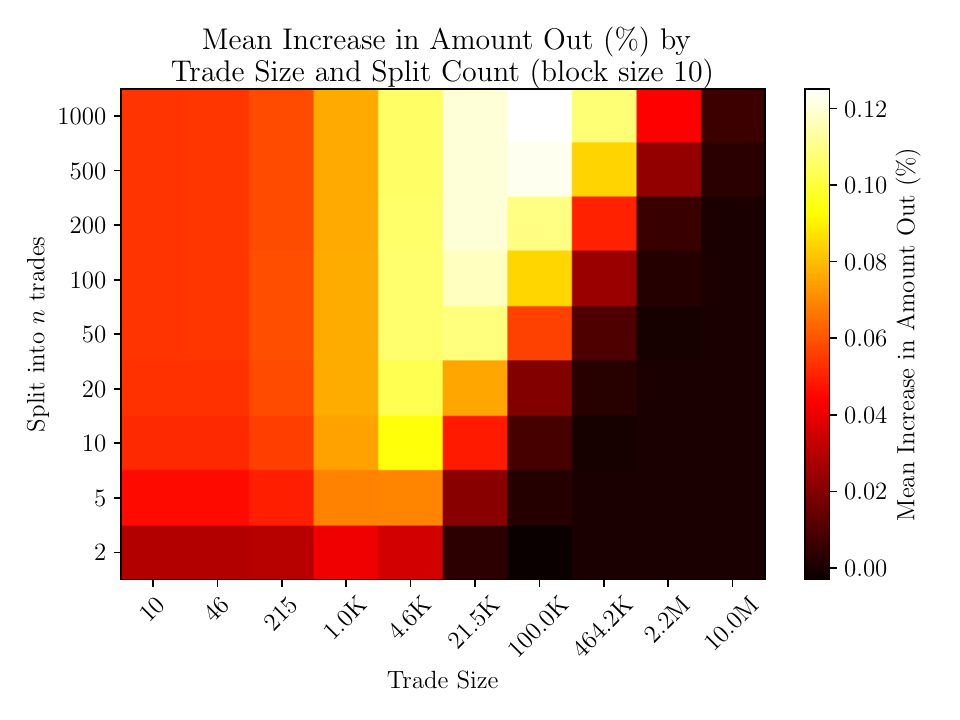}
  \end{minipage}\hfill
  \begin{minipage}{0.35\columnwidth}
    \caption{
      Trade-splitting experiment outcome. For various trade sizes, we compare the amount out (without splitting) vs the amount out after splitting the single trade $n$ equally-sized trades.
    }
    \label{fig:trade-split-heatmap}
  \end{minipage}
\end{figure}

The social planner may want to encourage order-splitting, especially when processing large transactions with significant price impact.\footnote{In an environment where network(gas) fees need to be paid for each transaction's execution, splitting the transactions may have boundaries that both traders and the social planner will need to take into account.}
We perform the following two experiments to test whether traders can benefit from splitting their transactions under \clvr.
First, we check if a trader can expect better execution (in expectation) when splitting the transaction into two or more.
We start by generating a `buy' trade of size $s$, which we will refer to as $t^*$.
Then, we generate 9 additional trades randomly, according to the log-normal distribution described above.
We sequence these 10 trades using \clvr, and compute the amount of token received by $t^*$.
Then, we split $t^*$ into $n$ equally-sized trades
$t^*_1, \cdots, t^*_n$,
and again sequence the trades and sum the amount of token received by each trade $t^*_i$.
This process is repeated 1,000 times for each combination of transaction size $s$ and number of splits $n$.
The results of this experiment are drawn in Figure~\ref{fig:trade-split-heatmap}.
We find that for all 10 trade sizes tested (varying from 10 to 10 million), the trader experiences a slight increase in amount out, on average, which is increasing in the number of splits $n$.

\begin{figure}[htb]
  \centering
  \begin{minipage}{0.6\columnwidth}
    \includegraphics[width=\columnwidth]{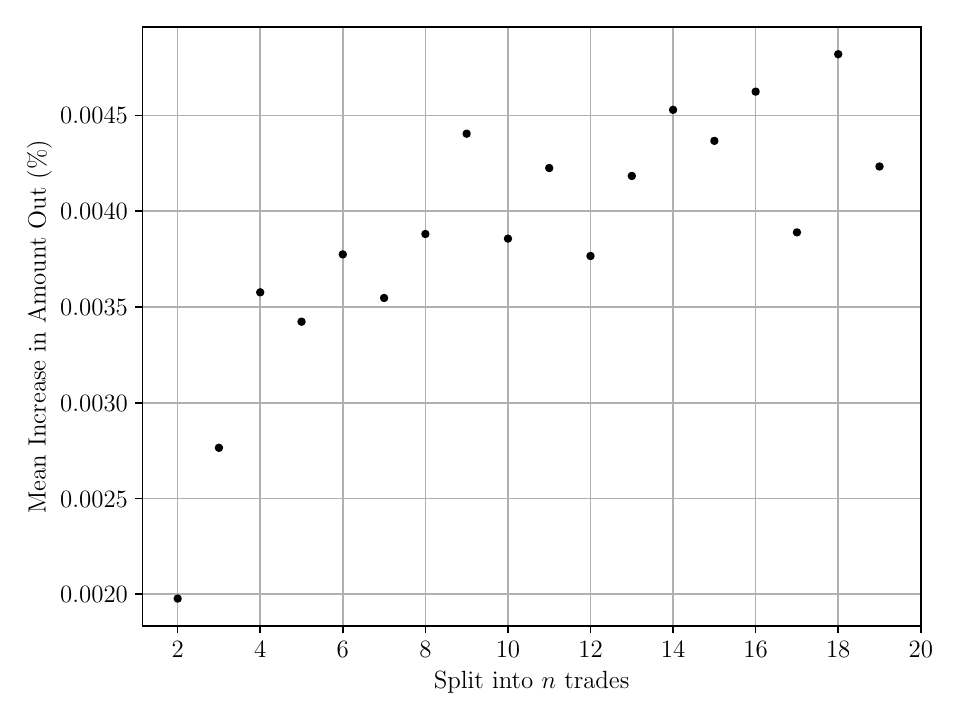}
  \end{minipage}\hfill
  \begin{minipage}{0.35\columnwidth}
    \caption{
      Trade-splitting experiment outcome (every trader splits). We provide evidence the mean percent increase in amount out when every trader splits into $n$ trades.
    }
    \label{fig:trade-split-everyone-splits}
  \end{minipage}
\end{figure}

Next, we examine what would happen to the ecosystem if \textit{all} traders split their transactions given the benefits that we show in Figure~\ref{fig:trade-split-heatmap}.
We perform the following experiment.
First, we generate 10 trades at random using the log-normal distribution described above.
Then, we sequence the trades using \clvr and compute the amount out for each trade.
Finally, we split each trade into $n$ equally-sized trades, sequence them using \clvr, and again compute the amount out that each original trader would receive from their several trades.
The process is repeated 1,000 times.
We plot the results in Figure~\ref{fig:trade-split-everyone-splits}.
We find an average increase in amount out, which increases in the number of splits $n$.
To conclude, if one trader splits her transaction (no matter the transaction size), she can get better execution (in expansion) by splitting the transaction.
Therefore, if all traders adopt this strategy, we show that they achieve better execution by splitting their transactions. This suggests that under \clvr, traders can benefit from splitting their transactions and thus have an incentive to do so.

\section{Computational Complexity of \clvr}
\label{appendix:computational_cost}

In the main body of this work we present \clvr as an algorithm with $O(n^2)$ complexity: a simple loop-within-loop construction to minimize volatility at each step.
In this section we describe a simple modification to achieve $O(n\cdot \log\ n)$ performance.

Intuitively, this modification relies upon the fact that at each step \clvr selects the next trade which minimizes an objective function: the single-step contribution to volatility.
In the main body of this work we present this selection process as a simple linear scan over the trades which are not yet sequenced.
Suppose, instead, that we begin by dividing the trades by their direction (buy / sell), and then sorting each list in a sorted collection keyed by the trade's size.
Then, at each step, we are able to perform a selection from each collection via binary-search.
The following contains a high-level sketch.

Consider an input list of trades and an AMM with a current price $p_0$.
Divide the trades into balanced binary trees \texttt{trades\_buy} and \texttt{trades\_sell}, keyed by their quantity of assets traded.
This consumes $O(\log\ n)$ computation in upfront preparation.
For each step $i$, repeat the following process.

Suppose at step $i$ the current price is $p_i$.
Compute the ideal trade direction and volume which would return the price to $p_0$ (the set-point) -- this is an $O(1)$ operation.
Intuitively, the best trade to sequence will have a quantity which is most close to the computed ideal.
We now can search for (at most) three candidate trades.
From the list of trades in the ideal direction, use the ideal quantity to lookup the nearest trades, both larger and smaller in quantity, if present.
This is $O(\log\ n)$.
From the list of trades in the opposite direction, lookup the smallest such possible trade, if present.
This is another $O(\log\ n)$.
Among these (at most) three candidates, locate the ideal trade and select it.
This is $O(1)$.
Finally, delete the trade from the balanced binary tree, this is $O(\log\ n)$.

Thus, we have reduced the inner loop's complexity to $O(\log\ n)$ for each of $n$ steps, and \clvr executes in $O(n\cdot \log\ n)$ time.